\documentstyle[prl,preprint,aps,psfig]{revtex}
\begin{document}
\draft
\title{Activation gaps for Fractional Quantum Hall Effect: Realistic 
Treatment of Transverse Thickness}
\author{K. Park, N. Meskini, and J. K. Jain}
\address{Department of Physics, 104 Davey Laboratory,
The Pennsylvania State University, University Park, Pennsylvania 16802}
\date{\today}
\maketitle
\begin{abstract}

The activation gaps for  
fractional quantum Hall states at filling fractions $\nu=n/(2n+1)$ are
computed for heterojunction, square quantum well, as well as parabolic 
quantum well geometries, using an interaction potential  
calculated from a self-consistent electronic
structure calculation in the local density approximation.
The finite thickness is estimated to make 
$\sim$30\% correction to the gap in the heterojunction geometry 
for typical parameters, which accounts for roughly half of the discrepancy
between the experiment and theoretical gaps computed for 
a pure two dimensional system.  
Certain model interactions are also considered.  It is
found that the activation energies behave qualitatively differently depending on
whether the interaction is of longer or shorter range than the Coulomb
interaction; 
there are indications that fractional Hall states 
close to the Fermi sea are destabilized for the latter.

\end{abstract}


\section{Introduction}

A fundamental aspect of the phenomenon of the 
fractional quantum Hall effect (FQHE) \cite {Tsui} is the  
existence of a gap at certain Landau level fillings
in the excitation spectrum for a disorder-free system, which 
is responsible for properties like fractional charge 
and the fractionally quantized Hall resistance \cite {Laughlin}.  An
understanding of the physical origin of the gap
lies at the heart of the FQHE problem. 

The composite fermion (CF) theory \cite {Jain,Review1,Review2} 
gives a simple intuitive explanation for the existence of the
gaps.  First, electrons capture an even number of vortices 
to become composite
fermions, since this is how they can
best screen the repulsive interaction.   
As a consequence of the phases produced by the vortices, composite
fermions experience a reduced effective magnetic field.  They form Landau
levels (LLs) in the reduced magnetic field, called CF-LLs in 
order to distinguish
them from the Landau levels of electrons.  A gap in the excitation spectrum
occurs whenever composite fermions fill an integer number of CF-LLs.
This provides an excellent description of the phenomenology of the  
FQHE; in particular, it gives a simple explanation for the observed fractions
at $\nu=n/(2pn\pm 1)$, which correspond simply to $n$ filled LLs of composite
fermions carrying $2p$ vortices.  Thus, an effectively single particle 
description of the strongly correlated electron liquid 
becomes possible in terms of composite fermions. 
The CF physics was spectacularly confirmed also in tests 
against exact results known for finite system from numerical diagonalization
studies.
The CF wave functions were found to have close to 100\% overlap with the exact
eigenfunctions, and predicted energies with an accuracy of
 better than 0.1\% for
systems of up to 12 particles \cite {Review1,tests}.  
A comparison with exact diagonalization results established that the 
gaps predicted by the CF theory are accurate to within a few percent.
However, it is only relatively  
recently that it has become possible to make more detailed 
quantitative comparisons between theory and experiment.  
The main hurdle was the lack of a suitable method for dealing with large
systems of composite fermions.  In recent years we have developed a
technique \cite {JK} that allows us to carry out 
Monte Carlo calculations on systems containing as many as 60 composite
fermions, which are sufficiently large to obtain reliable information on 
FQHE states at least up to 6/13. This paper reports on the results of 
our Monte Carlo calculations
for the gaps of various FQHE states, extending our 
previous work \cite {Park} as well as correcting some of the
assertions made therein.  The main new feature in this 
work is that we take account of the non-zero thickness of the electron 
system by determining the effective interaction in self-consistent 
local density approximation (LDA).  The calculations contain no adjustable
parameters; the only inputs are the shape of the confinement potential
(heterojunction, square quantum well, or parabolic quantum well)  
and the electron density.

There is a long history of calculation of gaps in the FQHE, dating back to 
the work of Laughlin \cite {Laughlin}.  
Accurate estimates for the gap of the FQHE state at $\nu=1/3$ for 
a strictly two dimensional (2D) system, where the interaction 
between electrons is of ``pure
Coulomb" form (as opposed to an ``effective" 
interaction after non-zero thickness is
taken into account), were obtained by Morf and Halperin \cite {Morf}
in a Monte Carlo calculation, using the variational wave functions 
of Laughlin, and by Haldane and Rezayi 
\cite {Haldane} from small-system, exact-diagonalization calculations.  
For lack of accurate wave 
functions, the gaps of other FQHE states could be estimated initially 
only from exact diagonalization calculations \cite {Fano}.
However, as one goes along the sequence 
$\nu=n/(2n+1)$, it takes larger and larger numbers of particles to get
reliable values for the gaps; since the size of Hilbert space increases
exponentially with $N$,  the exact diagonalization studies 
are of little use for large $n$.  For example, only two systems can be
studied at present for 3/7 (with 9 and 12 particles), and no exact
diagonalization is possible for 4/9, which requires at least 16 particles.
The gaps for the pure Coulomb interaction  
were computed by Jain and Kamilla \cite {JK} for several FQHE
states within the framework of the composite fermion theory, which 
we believe provide accurate estimates for an idealized 
zero-thickness system with no disorder.

It has been well known for quite some time that while 
the pure Coulomb gaps do give a rough estimate of the magnitude of the
experimental gaps, certain quantitatively significant 
effects present in real experiments 
must be incorporated for a more detailed comparison.
While these do not require any new conceptual input, it is important to
ascertain the relative importance of these effects, and to convince ourselves
that we are not missing any physics.  The aim of this work is to 
investigate one of these effects, namely the modification in the
interelectron interaction originating from
the finite transverse extent of the electron wave function,
in as much detail as is possible at the present.  
Since the early calculations of Zhang and Das Sarma (ZDS) \cite {ZDS} and 
Yoshioka \cite {Yoshioka}, much of the work 
dealing with the finite thickness has employed model interaction 
potentials, e.g. the ZDS potential $e^2/(\lambda^2+r^2)^{1/2}$,  
which  simulate the effect of non-zero thickness by 
softening the interaction at short distances.  The parameter characterizing
the thickness in these potentials must be  determined from
other considerations.  Very recently, Ortalano {\em et al.} \cite {Ortalano}
carried out a
calculation of the gap at 1/3 by feeding into their exact diagonalization
study the interaction that they obtain form a self-consistent LDA
calculation.  Park and Jain \cite {Park} computed the gaps of 
various other FQHE states
using the ZDS model, fixing the thickness parameter $\lambda$ 
by requiring that the gap for the 1/3 state agree with that obtained by
Ortalano {\em et al.}  This produced an excellent 
agreement between the theoretical
and experimental gaps.  However, a comparison with the more realistic
Stern-Howard
\cite{Howard,Ando} interaction led  Morf \cite {Morf99} 
to conclude that the value of the $\lambda$ used in this work 
was too large by approximately a factor of two, and therefore the 
comparison with experiment was not valid and the agreement fortuitous.  
To resolve this issue, and also to obtain reliable values for the gaps, 
we have computed the gaps directly from  
the interaction obtained from the self-consistent LDA.  It is found that 
the non-zero thickness makes 20-50\% correction for typical
experimental parameters, reducing 
the discrepancy between pure 2D 
theory and experiment approximately by half,   
but the theoretical gaps still significantly overestimate the gaps \cite
{Park2}.

There are other effects that will diminish the gaps beyond their values
obtained in the present work.  One assumption here is that the electronic
states are confined to the lowest LL, which is indeed a valid approximation
in the limit of sufficiently large magnetic fields, but, at typical 
experimental fields, Landau level mixing may not be negligible.
It is expected that the CF particle and hole excitations will 
lower their energies by an admixture with
higher Landau levels.  Previous estimates \cite {LLmixing}
suggest that it is roughly a 20\%
effect.  The omnipresent disorder, neglected in the present study,
is also expected to
reduce the gaps.  A reliable theoretical treatment of these issues
is beyond the scope of the present work.

We also calculate gaps for different kinds of model potentials, 
some differing from the Coulomb interaction 
in the short-distance behavior while others in the long range.
The qualitative behaviors give an indication  
of a relation between the range of the potential and the 
stability of the CF sea.

The paper is organized as follows.  In Section II, we give a brief account of
the computational methods, mentioning, in particular, certain modifications in
the self-consistent LDA in this work,  
appropriate for the problem at hand; 
a summary of the CF wave functions and the Monte Carlo is also given.  
Section III gives our results for gaps for various densities in three sample
geometries:  heterojunction, square quantum well, and parabolic quantum
well.  Section IV discusses a comparison of our results with experiment,
Section V contains gaps for several model interactions, 
and the paper is concluded in Section VI.
 
\section{Computational details} 

For completeness, we provide a  brief outline of our computational methods.  
Readers interested in further details can find them in the literature. 

{\bf Self Consistent LDA}

Following the standard approach \cite {Ando,Stern}, one solves 
self-consistently  
the one dimensional Schr\"odinger and Poisson equations for the
direction perpendicular to the plane of the 2D electron system 
(taken along the $z$ axis here):

\begin{equation}
\left(-\frac{\hbar^2}{2m} \frac{d^2\;\;}{dz^2}+V_{eff}(z)\right) \xi(z) 
=E\xi(z)
\end{equation}

\begin{equation}
V_{eff}(z)=V_W(z)+V_H(z)+V_{XC}(z)
\end{equation}

\begin{equation}
\frac{d^2 V_H(z)}{dz^2}=-\frac{4\pi e^2}{\epsilon}[\rho(z)-\rho_I(z)]
\end{equation}

\begin{equation}
\rho(z)=N|\xi(z)|^2
\end{equation} 
Here, $V_H$, $V_W$ and $V_{XC}$ are the Hartree, confinement, and
exchange-correlation potentials, and $\rho_I$ is the density of the 
ionized donor atoms.  The exchange correlation potential is
assumed to depend only on the local density, which is usually a
quite reasonable approximation. 

The equations above have been slightly modified from the ones used at zero
magnetic field \cite
{Ortalano} to suit our present purpose.
In the past, it has been assumed that 
the effective potential was not significantly affected by
application of the magnetic field, and the zero field effective 
interaction was used for  
high magnetic fields as well \cite {Ortalano}.
We make a few changes from the
standard zero field calculation, which we believe are appropriate when 
discussing electrons confined to the lowest Landau level.
These are as follows:

(i) At zero magnetic field, electrons occupy one or several subbands 
depending on the density.  In our calculations below, we  
assume that they occupy only one subband.  This would 
clearly be unphysical for sufficiently large densities 
at zero magnetic field, but is appropriate at large
magnetic fields, where only the lowest LL is occupied.
This makes no difference at low densities, 
when the self consistent solution at zero field also involves only one
subband, but the results are affected somewhat at large densities.  
We have not investigated how much of a quantitative difference this alteration
makes at high densities.

(ii)  We assume that electrons are fully polarized.  This of course 
is motivated by the fact that we are interested in fully polarized electronic
states, which is appropriate for  
sub-unity filling factors at high magnetic fields.

We further make the following approximations.

(iii) For the exchange correlation energy 
we use the form given by Vosko {\em et al.} \cite {HB} 
rather than the more usual one by Hedin and
Lundqvist \cite {Hedin}, the former being more appropriate for 
spin polarized electrons.
This does not make appreciable quantitative difference; in fact, leaving out
the exchange correlation corrections entirely is also a rather good
approximation for the present problem.

(iv) In heterojunction, the electron wave function has a small 
amplitude on the AlGaAs side, with most of the wave function being 
confined to
the GaAs side.  A proper treatment will require 
taking a position dependent dielectric function
as well as a position dependent 
mass, and replacing the step function change
in these quantities at the interface 
by a smooth function in order to ensure
that the calculations are technically well controlled \cite {Stern}.  
In order to avoid these complications,
we have assumed that the wave function is entirely confined 
on one side of the
junction;  this was found to be an excellent approximation 
in earlier calculations \cite {Stern}.
Similarly, for the quantum well potential, 
we have assumed infinite barriers, keeping 
electrons out of the insulator. 
This should be a reasonable approximation provided the electron energies are
deep in the well.

(v) The depletion charge density in experimental samples  
is unknown, but often small \cite {Willett};
we have set it to zero in our calculations.  This may not be a good 
approximation especially when the electron density becomes comparable to the
depletion layer charge density.  Image charge effects due to a slight 
mismatch of the dielectric function at the interface have also been 
neglected. 

\begin{figure}
\centerline{\psfig{figure=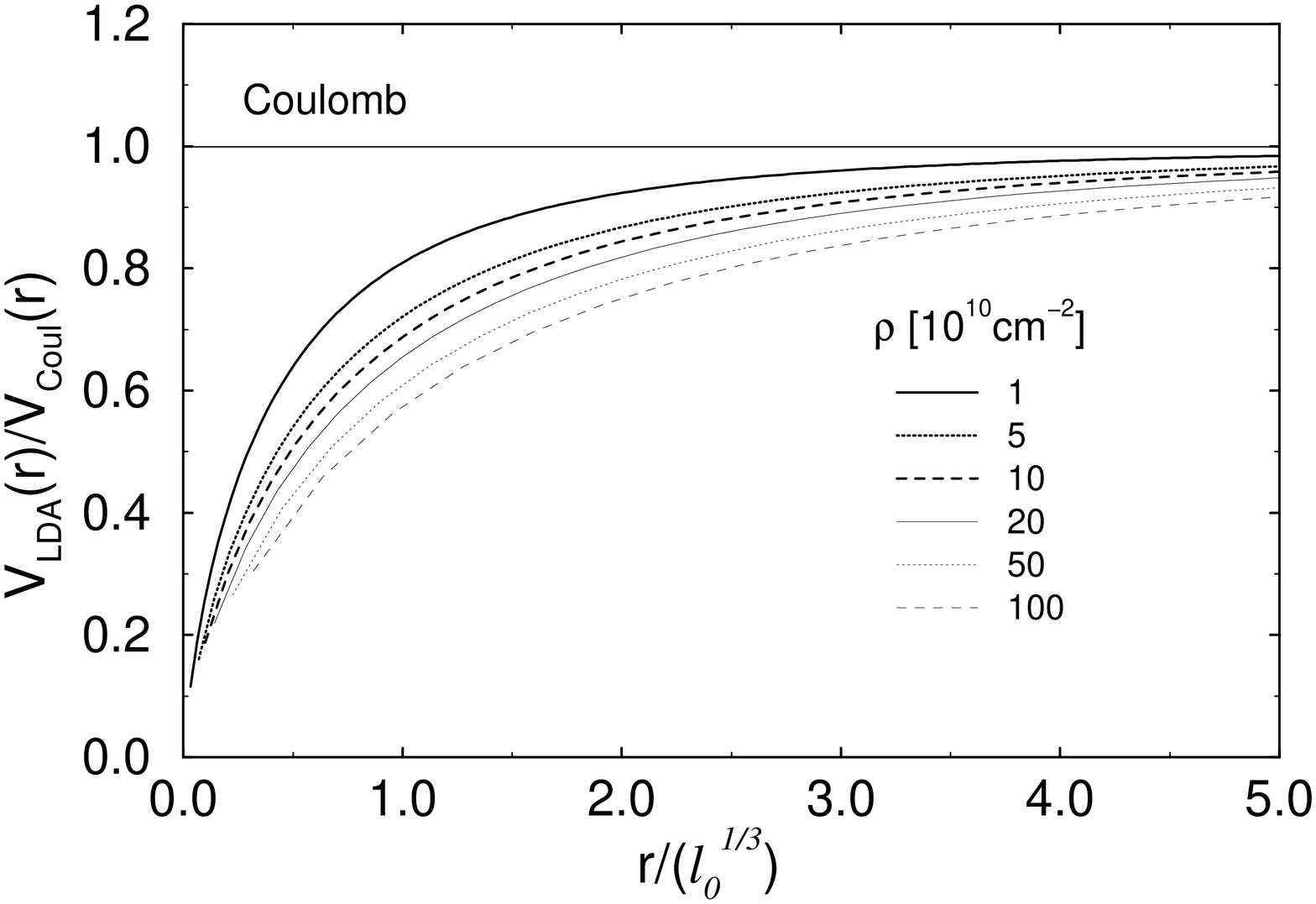,width=7.0in,angle=-90}}
\caption{The effective interaction, $V_{LDA}(r)$ for the heterojunction
geometry for densities ranging from $1.0\times 10^{10}$ cm$^{-2}$ 
to $1.0\times 10^{12}$
cm$^{-2}$.  The interaction is shown in units of the Coulomb interaction and
the distance is given in units of the magnetic length at $\nu=1/3$, $l_0^{1/3}$.
\label{fig1}}
\end{figure}

The above equations are solved by an iterative procedure until convergence is
obtained for $\xi(z)$.  The effective interaction potential is then given by 
\begin{equation}
V_{LDA}(r)=\frac{e^2}{\epsilon}\int dz_1 \int dz_2 \frac{|\xi(z_1)|^2 
|\xi(z_2)|^2}{[r^2+(z_1-z_2)^2]^{1/2}}
\end{equation}
The LDA interaction for the heterojunction geometry is shown in
Fig.~\ref{fig1}.

{\bf Composite Fermion Wave Functions}

We compute the energy gaps  
by evaluating the expectation 
values of the effective interaction energy 
$V=\sum_{j<k}V_{LDA}(r_{jk})$ in the 
composite fermion wave functions for the
ground and excited states:
\begin{equation}
\Delta =\frac{<\Phi^{CF-ex}|V|\Phi^{CF-ex}>}{<\Phi^{CF-ex}|\Phi^{CF-ex}>}
-\frac{<\Phi^{CF-gr}|V|\Phi^{CF-gr}>}{<\Phi^{CF-gr}|\Phi^{CF-gr}>}
\end{equation}  
where $\Phi^{CF-ex}$ and $\Phi^{CF-gr}$ are the CF wave functions for the
excited and the ground states, respectively.

We use the spherical geometry in this work, which
considers $N$ electrons on the surface of a sphere, moving
under the influence of a strong radial magnetic field.
Fully spin polarized electrons and a complete lack of
disorder are assumed. The flux through
the surface of the sphere is defined to be $2Q\phi_0$, where
$\phi_0=hc/e$ is the flux quantum and $2Q=$ integer.  The single
particle
eigenstates are the monopole harmonics \cite {Wu}, denoted by
$Y_{Q,n,m}(\Omega)$,
where $n=0,1,...$ is the LL index, $m=-Q-n, -Q-n+1, ...
Q+n$ labels the $2Q+2n+1$ degenerate states in the
$n$th LL, and $\Omega$ represents the angular coordinates
$\theta$ and $\phi$.  

According to the CF theory \cite {Jain}, the  problem of interacting
electrons at $Q$ is equivalent to that of weakly interacting composite
fermions at effective monopole strength $q=Q-p(N-1)$.
The many body
CF states can be constructed from the following
`single-CF' wave functions \cite {JK}:
\begin{eqnarray}
Y^{CF}_{q,n,m}(\Omega_j)= \tilde{Y}_{q,n,m}(\Omega_j)
\prod_{k}^{'} (u_j v_k-v_j u_k)^p\;,
\end{eqnarray}
\begin{eqnarray}
&&\tilde{Y}_{q,n,m}(\Omega_j) =
N_{qnm} (-1)^{q+n-m} \frac{(2S+1)!}{(2S+n+1)!}
u_j^{q+m}   v_j^{q-m}  \nonumber \\
&&
\sum_{s=0}^{n}(-1)^s {{n \choose s}} {{ 2q+n \choose q+n-m-s}}
\; u_j^s \;    v_j^{n-s}  \; {\bf U}_j^s \; {\bf V}_j^{n-s}\; ,
\end{eqnarray}
\begin{equation}
N^2_{qnm}=\frac{(2q+2n+1)}{4\pi}\frac{(q+n-m)!(q+n+m)!}{n!(2q+n)!}
\;\;,
\end{equation}
\begin{equation}
{\bf U}_j=p \sum_{k}^{'}\frac{ v_k}{
u_j v_k - v_j u_k} + \frac{\partial}{\partial u_j}\;\;,
\end{equation}
\begin{equation}
{\bf V}_j=p\sum_{k}^{'}\frac{
-  u_k}{u_j v_k - v_j u_k} + \frac{\partial}{\partial v_j}\;.
\end{equation}
Here the prime denotes the condition $k\neq j$,
$p$ is an integer, $S=q+p(N-1)/2$,
and the spinor coordinates are defined as \cite{Haldane2}
$u_j\equiv \cos(\theta_j/2)\exp(-i\phi_j/2)
$ and $v_j\equiv \sin(\theta_j/2)\exp(i\phi_j/2)$.
The binomial coefficient ${{\alpha \choose \beta}}$ is to be
set to zero if $\beta > \alpha$ or $\beta < 0$.
The subscript $n$ in $Y^{CF}_{q,n,m}$ labels the CF-LL index.
Note that the wave function of the $j$th composite fermion involves
the coordinates of {\em all} electrons.
In the form written above, 
the CF wave function is fully confined to the lowest electronic LL.

The wave functions for the system of many composite
fermions are the same as the corresponding wave functions of non-interacting
electrons at $q$, but with $Y_{q,n,m}$ replaced
by $Y^{CF}_{q,n,m}$. The incompressible ground state consists of
an integer number of filled LLs of composite fermions.
The excited states are constructed by promoting one composite fermion from
the topmost occupied CF-LL to the lowest unoccupied CF-LL, which creates a CF
particle hole pair.  We are interested in the energy of this excitation in 
limit that the distance between
the CF particle and the CF hole is very large,
so we consider the excited 
state in which they are on the opposite poles of the
sphere.  Prior to an extrapolation of 
our results to the limit of $N\rightarrow \infty$, we
correct for the interaction between the CF particle and the CF hole, which
amounts to a subtraction of $-(2p+1)^{-2}/2\epsilon\sqrt{Q}l_0$, the 
interaction energy for  
two point-like particles of charges $e/(2p+1)$ and $-e/(2p+1)$ at a distance
$2R$, where $R=\sqrt{Q}l_0$ is the radius of the sphere.   We also correct for
a finite size deviation of the 
density from its thermodynamic value, by multiplying by a factor
$\sqrt{\rho/\rho_N}=\sqrt{2Q\nu/N}$, where $\rho$ is the thermodynamic density
and $\rho_N$ is the density of the $N$ particle system.

We emphasize here that
both the ground and excited state wave functions above
{\em contain no adjustable parameters}; they are completely
determined by symmetry in the restricted Hilbert
space of the CF wave functions.
Also, since the wave functions constructed here are strictly
within the lowest {\em electronic} LL, their energies will
provide strict variational bounds.  Of course, there is no variational
theorem for the energy {\em difference}, but the CF wave functions 
are known to be extremely accurate; they produce gaps with an
accuracy of a few percent for a given interaction potential, at least for
1/3, 2/5, and 3/7, for which exact results are known for finite systems.  
For these fractions, any error in the gaps will owe its origin mainly  
to various approximations in our calculation
of the effective interaction; 
insofar as finite width effects are concerned, 
we expect our gap calculations to reliable at the level of 
20\% \cite{Ortalano}.

An unusual feature 
of the composite fermion wave functions is that they are independent of the
actual form of the interaction, since they have 
no parameters to adjust.  While this may seem objectionable at first sight,
it captures the fact that the actual wave
functions (as obtained, say, in exact diagonalization studies)
are also largely insensitive to the form of
the interaction.  This rigidity to perturbations 
can be understood physically by analogy to the integer QHE.
The electron wave functions at integer
filling factors are quite independent of the interaction 
provided that it is
small compared to the cyclotron gap (i.e., LL mixing is negligible).  
In the CF theory, this would imply that
the interaction dependence of the wave function is negligible so long as the
residual interaction between the composite fermions is weak compared 
to the CF cyclotron gap.
For the $n/(2n+1)$ states with large $n$, the CF cyclotron gap may not
necessarily be large compared to the inter-CF interactions, 
and the actual wave functions may have some dependence on the form of the
interaction. While 
the actual wave functions are not known for these states, it may be
possible to investigate the issue in a variational approach by 
incorporating some variational degree of freedom which allows 
mixing between {\em CF} LLs to determine the extent to which the CF wave
function is perturbed.  We will, however, continue to work with the 
unperturbed composite fermion wave functions here, with the caveat that the
results may not be completely reliable at large $n$ (we suspect though, that
the intrinsic error in the CF wave function may still be small compared to the
uncertainty arising from Monte Carlo and from various approximations 
involved in evaluation of the effective interaction).

{\bf Monte Carlo}

The Monte Carlo employed in our work 
is quite standard.  Unfortunately, it is not possible to use 
in our problem certain clever time-saving 
techniques for updating fermion Slater determinants \cite {Ceperley}, 
since moving a single
particle alters all elements of the determinant, due to the strongly
correlated nature of the problem (remember, the wave function of each 
composite
fermion depends also on the positions of all other composite fermions).   
Therefore, we must compute the full Slater determinant at each step, which 
takes $O(N^3)$ operations rather than $O(N^2)$.
However, we are able to improve on the accuracy by moving 
all particles at each step.  
We note that the ground and excited state 
energies must be evaluated extremely accurately in order to get a reasonable
estimate for the gap, which is an $O(1)$ quantity. 
We also utilize the fact that the ratio of the gap to a reference
gap (say, for the pure Coulomb interaction) has much smaller variance 
than the gap itself from one Monte Carlo run to another.
A typical calculation of the energy gap requires 10$^7$ Monte Carlo steps,
taking up to 200 hours of computer time on a 500 MHz workstation.

Since there are no edges in the geometry being studied, we 
expect that the gap will have a linear dependence  on $N^{-1}$ to leading
order, which is also borne out by our results.
Therefore, we obtain the thermodynamic limit by a
linear fit to the finite system gaps.  The error is determined by the standard
least square method. 

\section{Results}

\begin{figure}
\centerline{\psfig{figure=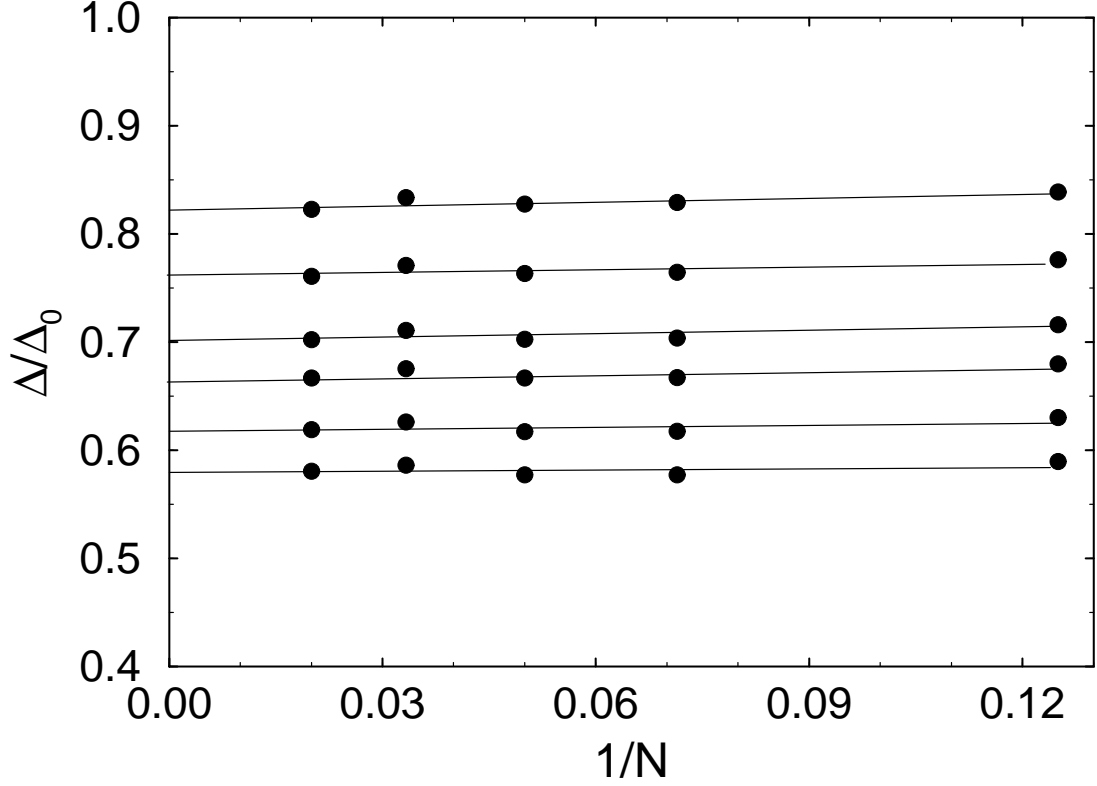,width=7.0in,angle=-90}}
\caption{Extrapolation of the activation gap at $\nu=2/5$ to the 
thermodynamic ($N^{-1}\rightarrow 0$) limit for the heterojunction geometry 
for densities (starting from top) of 
$1.0\times 10^{10}$ cm$^{-2}$, $3.0\times 10^{10}$ cm$^{-2}$,
$1.0\times 10^{11}$ cm$^{-2}$, $2.0\times 10^{11}$ cm$^{-2}$, 
$5.0\times 10^{11}$ cm$^{-2}$,
and  $1.0\times 10^{12}$ cm$^{-2}$.  The Monte Carlo uncertainty is smaller
than the symbol size and the solid line is the best straight line fit.
Systems of up to 50 composite fermions
\label{fig2}}
\end{figure}

\begin{figure}
\centerline{\psfig{figure=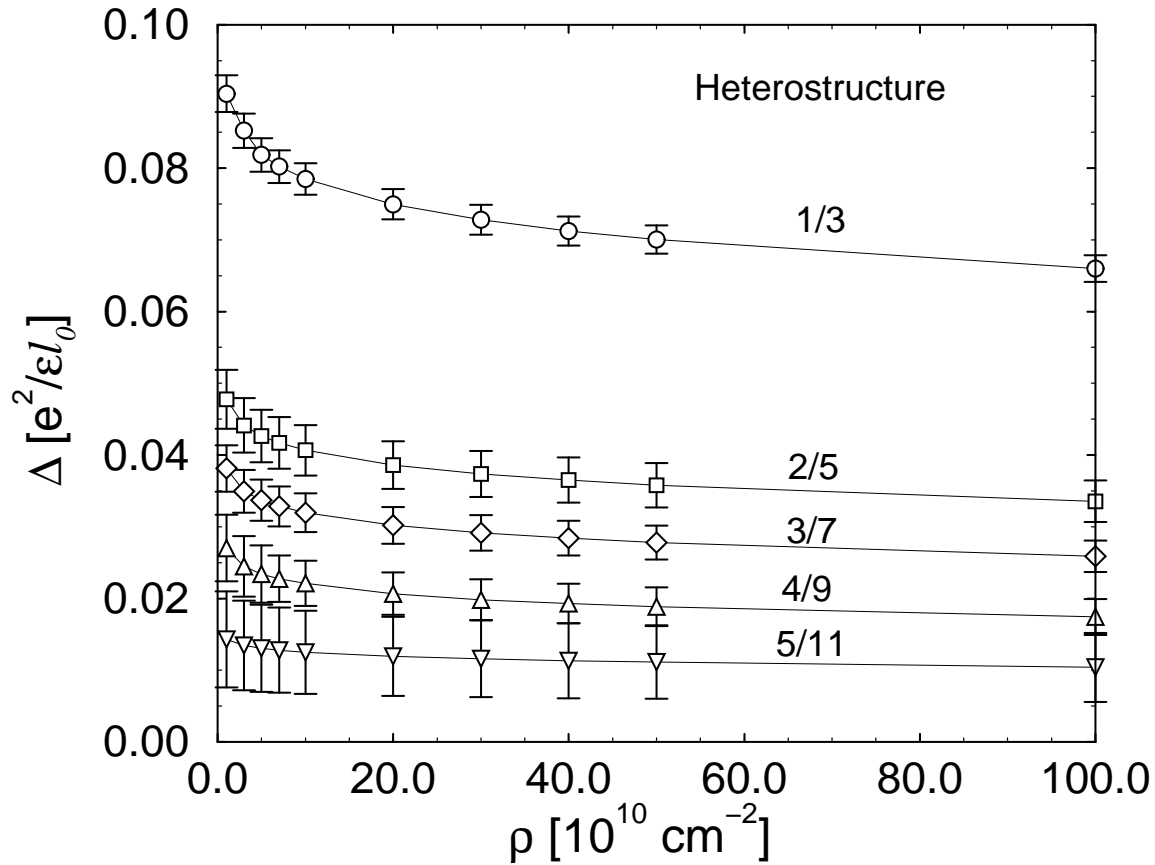,width=7.0in,angle=-90}}
\caption{The CF predictions for the gaps in heterojunction geometry as  
a function of the density ($\rho$) 
ranging from $1.0\times 10^{10}$ cm$^{-2}$ to 
$1.0\times 10^{12}$ cm$^{-2}$, for FQHE states at 1/3, 2/5, 3/7, 4/9, and
5/11, with the filling factors indicated on the figure.  The gaps are
expressed in $e^2/\epsilon l_0$ where $\epsilon$ is the dielectric constant of
the background material ($\epsilon \approx 13$ for GaAs) and $l_0$ is the
magnetic length.  
\label{fig3}}
\end{figure}

\begin{figure}
\centerline{\psfig{figure=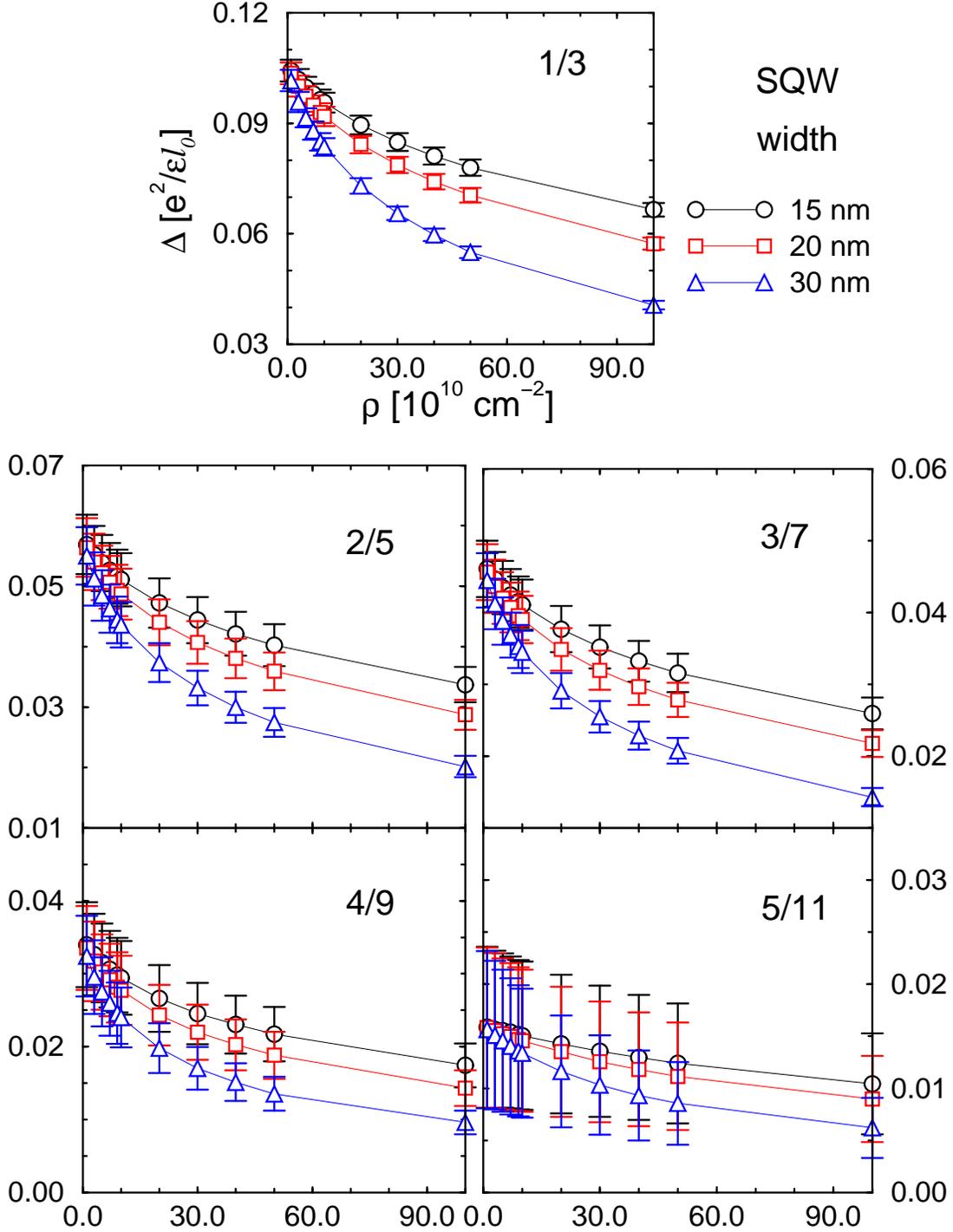,width=6.0in,angle=0}}
\caption{The CF predictions for the gaps in square quantum well (SQW) 
geometry for
densities ranging from $1.0\times 10^{10}$ cm$^{-2}$ to 
$1.0\times 10^{12}$ cm${-2}$ for
quantum well widths of 150 A$^0$, 200 A$^0$, and 300 A$^0$.
The filling factors are shown on the figure.
The labels for axes are shown only for 1/3 for convenience.
\label{fig4}}
\end{figure}

\begin{figure}
\centerline{\psfig{figure=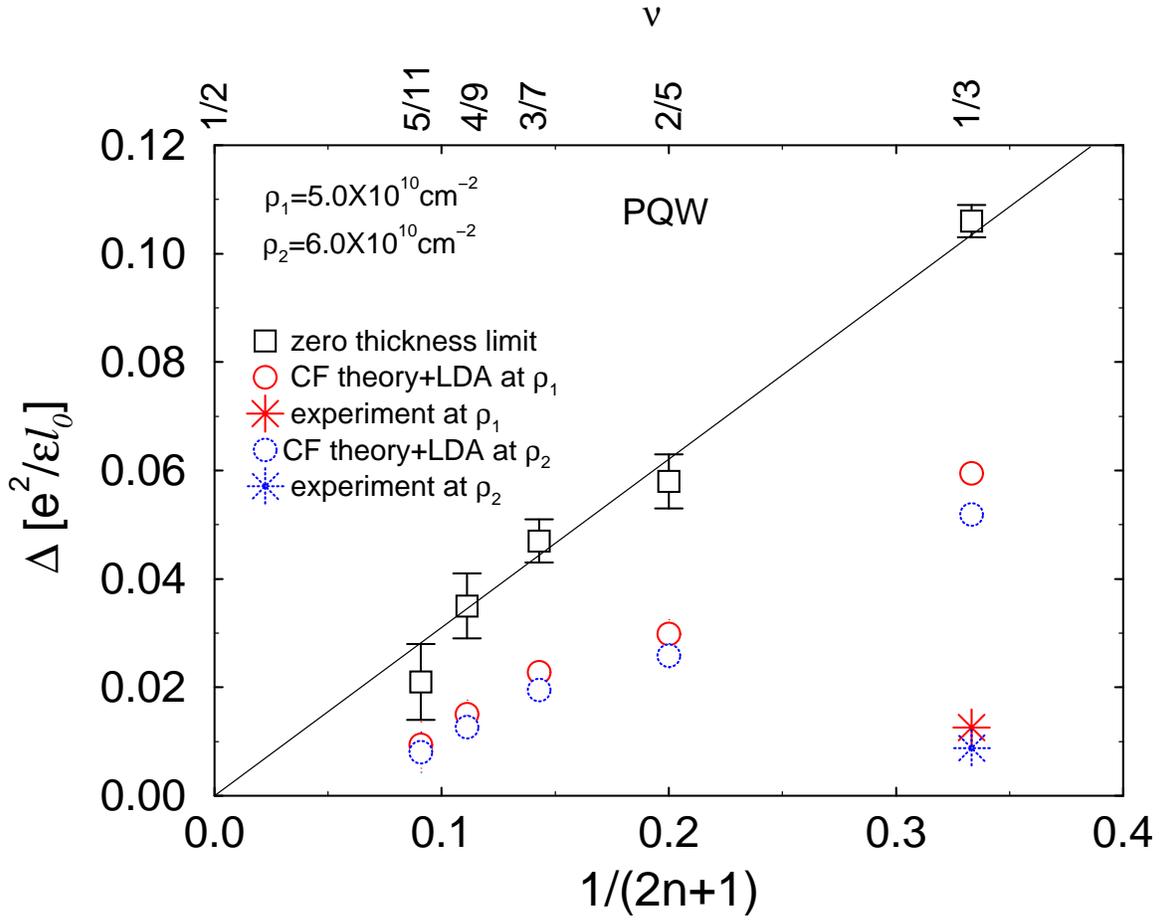,width=7.0in,angle=-90}}
\caption{The CF predictions for the gaps in parabolic quantum well 
(PQW) geometry for two densities $5.0\times 10^{10}$ 
cm$^{-2}$ and $6.0\times 10^{10}$
cm$^{-2}$ as a function of filling factor (top axis).  
The squares are 
for pure Coulomb interaction, circles for the LDA interaction, and stars
are the experimental results taken from Shayegan et al. 
\protect \cite {Shayegan}.  
The experimental PQW is 3000 A$^0$ wide, with curvature $\alpha = 5.33\times
10^{-5}$ meV/A$^{0-2}$ and barrier height from the bottom $V_0=276$ meV.  
We have set the barrier height to infinity in our LDA calculations.
\label{fig5}}
\end{figure}

The only inputs in our calculations are the electron density and 
the sample geometry. We have computed the gaps for a range of 
densities (from $1.0 \times 10^{10}$ cm$^{-2}$ to
$1.0 \times 10^{12}$ cm$^{-2}$) and for three sample
geometries most popular in experiments: single heterojunction, square quantum  
well (SQW), and parabolic quantum well (PQW).
All results have been obtained by an extrapolation of the finite system
results to the limit $N^{-1}\rightarrow 0$, as shown for the case of
$\nu=2/5$ in Fig.~\ref{fig2};  systems of up to 50 particles were considered
for the extrapolation.  Figs.~\ref{fig3}, \ref{fig4},  
and \ref{fig5} show the gaps as a function of density for 
several sample geometries.  Since the ratios $\Delta/\Delta_0$ are determined
quite accurately, as seen in Fig.~\ref{fig2}, the
uncertainty in $\Delta$ comes almost entirely from  
$\Delta_0$, for which we use values given in Ref. \cite {JK}.
For a typical sample
density of $2\times 10^{11}$ cm$^{-2}$, the 1/3 gap is reduced roughly 
by 30\% in a heterojunction, by 30\% in a square quantum well
of width 300$A^0$, and by 50\% in parabolic quantum well.
As expected, the gaps approach their Coulomb values
at small densities in the heterojunction geometry, and also at small 
QW widths in the quantum well geometries.  

A similar calculation for the gap was carried out by Ortalano {\em et al.}
for $\nu=1/3$, who obtain a bigger gap reduction,  
for reasons that are not known at the moment.  The 
pseudopotentials from our effective interaction are in agreement with 
theirs, provided the Bohr radius is set equal to the
magnetic length.  The gaps reported in Ref.\cite {Ortalano}  were for a six
particle system whereas we have determined the thermodynamic limit,
which may account 
for part of the discrepancy;  also, the result of Ref.
\cite{Ortalano}  was obtained from an exact
diagonalization of the Hamiltonian 
as opposed to our calculations which employs the 
CF wave functions,  but this ought not
to cause more than a few \% difference.

\section{Comparison with experiment}

\begin{figure}
\centerline{\psfig{figure=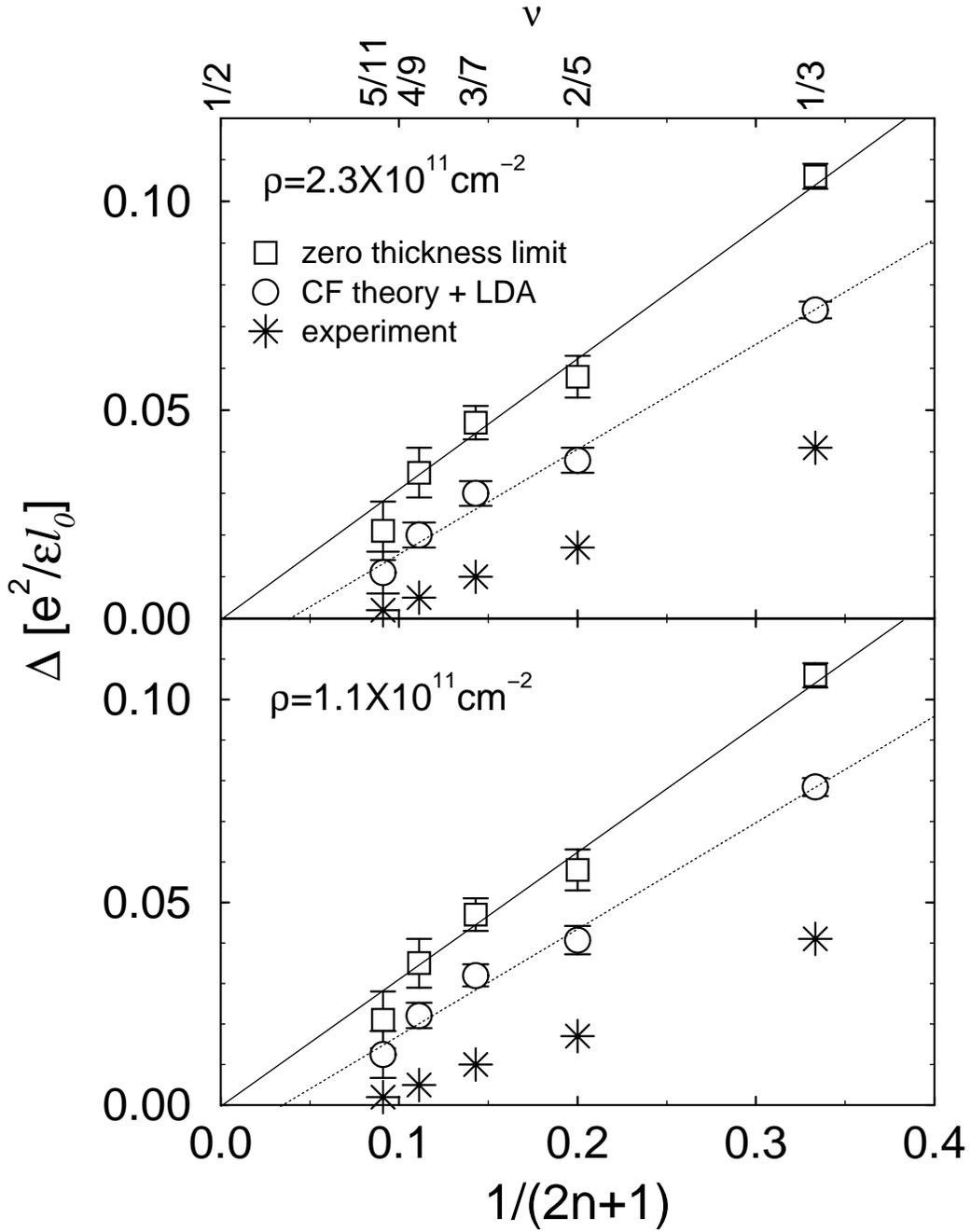,width=6.0in,angle=0}}
\caption{Comparison of the theoretical and the experimental gaps for the
heterojunction geometry for two different densities shown on the figures.
The squares are for pure Coulomb interaction, circles 
for the LDA interaction, and stars are taken from the experiment of Du
{\em et al.} \protect \cite {Du} 
\label{fig6}}
\end{figure}

\begin{figure}
\centerline{\psfig{figure=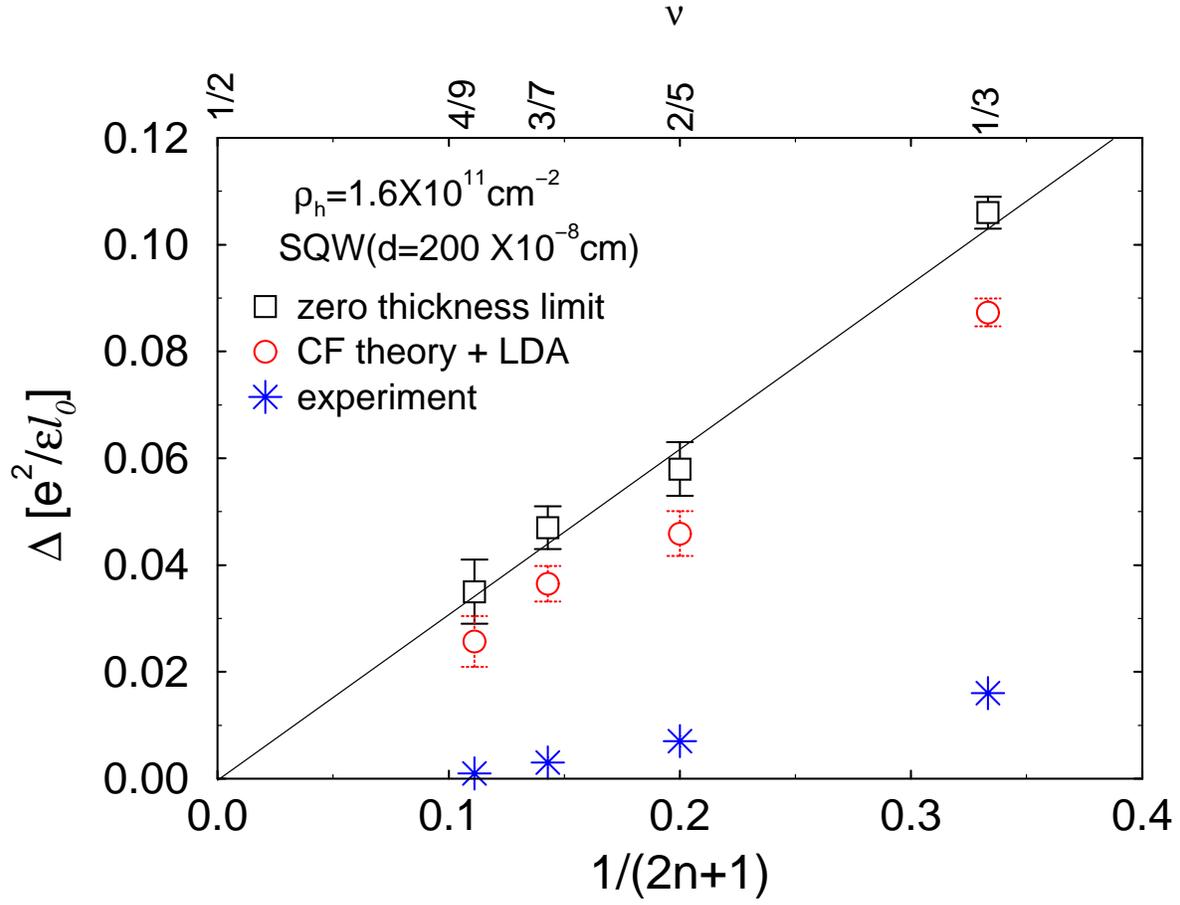,width=7.0in,angle=-90}}
\caption{Comparison of the theoretical and the experimental gaps for the
the square quantum well geometry.
The squares are for pure Coulomb interaction, circles 
for the LDA interaction, and stars are taken from Manoharan {\em et al.}
\protect \cite {Manoharan} 
\label{fig7}}
\end{figure}

Fig.~\ref{fig6} shows a comparison of our results for 
the heterojunction geometry with experiment on two densities \cite{Du}.
The finite thickness reduces the gaps from
their pure Coulomb values bringing them in better agreement with 
experiment.  Figs.~\ref{fig7} and \ref{fig5} compare our theoretical 
gaps with 
experimental gaps in square \cite {Manoharan} and parabolic 
\cite {Shayegan} quantum wells.  Here, again the 
gaps are reduced from their pure Coulomb values, but 
a substantial deviation still remains between theory and experiment. 

There are many possible 
sources that can cause disagreement between our theoretical gaps and 
the experimental gaps.  There are approximations involved in our
determination of the  effective interaction, which may
lead to a 20\% uncertainty in the theoretical gap values \cite {Ortalano}.
Then there are 
effects left out in the theory, namely Landau level mixing and disorder.   
Landau level mixing is likely to be most significant in the hole-type
samples (the square quantum well here \cite {Manoharan}), 
due to the relatively small cyclotron energy of holes.
The disorder is most relevant perhaps in parabolic quantum wells, 
due to alloy disorder,  which leads to 
relatively low mobilities; the strong suppression of the PQW gaps relative to
the computed values indicates that disorder can be rather important
quantitatively.  In view of this discussion, the 
comparisons of our results with heterojunction gaps are most meaningful.
One message one can take from these comparisons is that 
Landau level mixing and disorder also make a sizable correction to 
the excitation gaps in typical experiments. 
As mentioned earlier, for 5/11 and 6/13, the intrinsic errors in the
``unperturbed" CF wave functions, not yet quantified, may 
also be partly responsible for the deviations between theory and experiment.

An extrapolation of the experimental gaps suggests that they might  
vanish at a finite $n$.  Surely, any finite amount of disorder will  
cause such a behavior.  However, it is
an interesting question whether the gaps will vanish at a finite $n$ even in
the absence of disorder.  There is no fundamental reason that this could not
happen.  In our computations, 
while the Coulomb gaps extrapolate to zero at
$\nu=1/2$, within numerical uncertainty, the non-zero thickness gaps 
appear to vanish at a finite $n$  [along the sequence $\nu=n/(2n+1)$], at
least for a straight line fit through them.
This is clearest for relatively larger gap reductions, e.g. in the
heterojunction or the parabolic quantum well systems.
These results might indeed be
indicating an intrinsic absence of FQHE for $n$ larger than a critical value,
even for an ideal situation with no Landau level mixing and no disorder.
This does not imply, however, that the composite fermion theory becomes
invalid here, but only that composite fermions 
do not show integer QHE (IQHE), presumably because the residual
inter-CF interactions become increasingly 
significant as the gap decreases, finally
destroying the gap altogether.
(We note that for small magnetic fields, the electron system also does not 
exhibit IQHE; a better starting point here is the Fermi sea, with the
magnetic field treated as a perturbation, rather than 
a filled Landau level state.)
This kind of breakdown of the FQHE, if one actually occurs,  
will be due to a short-range modification in the
inter-electron interaction due to non-zero thickness, to be distinguished from
another possibility, discussed in the following section, 
which has to do with the {\em long-range} behavior of the interaction \cite
{HLR}.

\begin{figure}
\centerline{\psfig{figure=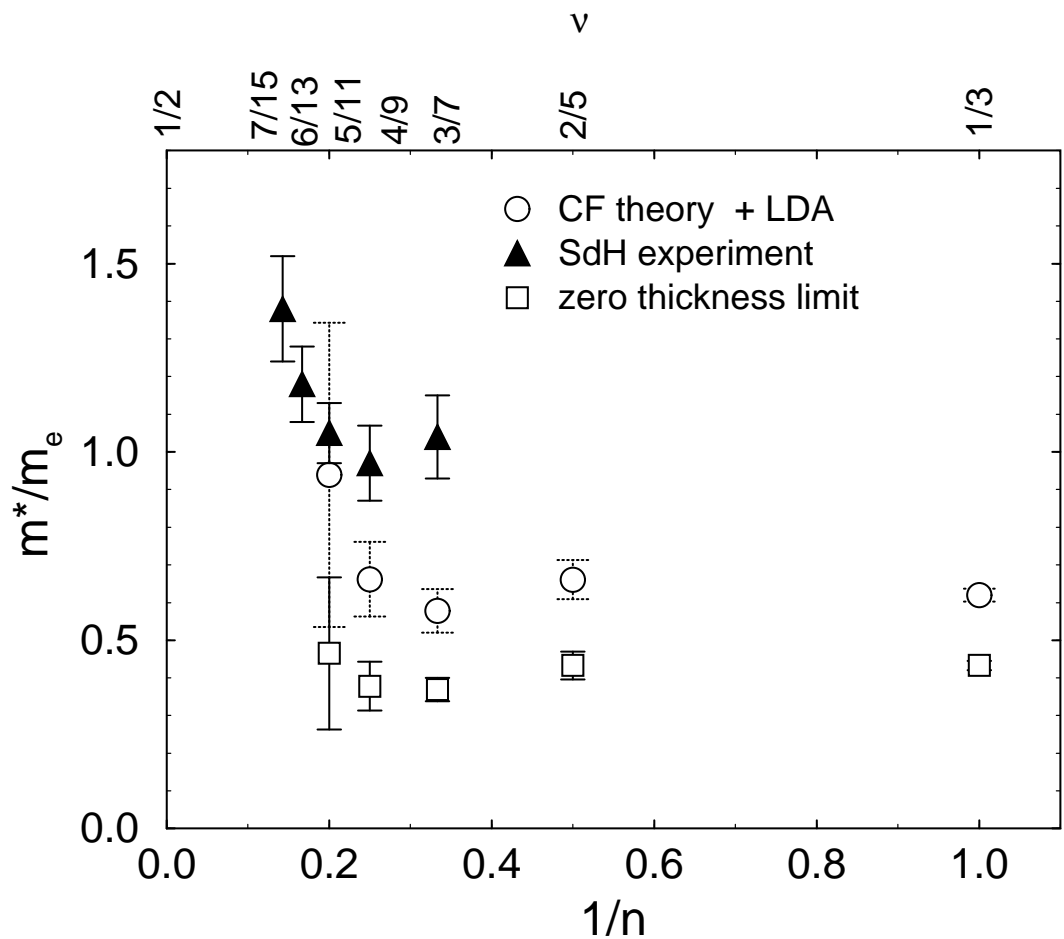,width=7.0in,angle=-90}}
\caption{The mass of composite fermion ($m^*$)
in units of the mass
of electron in vacuum ($m_e$) as a function of the filling factor for a
heterojunction sample with density $2.3 \times 10^{11}$ cm${^-2}$..
Both the mass computed from the theoretical
gaps in Fig.~\ref{fig6} (circles for the realistic calculation, squares
for zero transverse thickness) and that deduced from
an analysis of the SdH experiment (triangles, from
Du \protect {\em et al.} \protect \cite {DuSdH}) are shown.
\label{fig8}}
\end{figure}

The activation gaps can be equated to an effective cyclotron energy to define
an effective mass for the composite fermions \cite {HLR}:
\begin{equation}
\Delta =\hbar \frac{eB^*}{m^*c}=\frac{\hbar^2}{m^* l_0^2} \frac{1}{(2n+1)}\;
\end{equation}
where we have used that the effective field for composite fermions is given 
by $B^*=B/(2n+1)$ at $\nu=n/(2n+1)$.
On the other hand, since the gaps are determined entirely 
by the Coulomb interaction
(the only energy in the lowest LL constrained problem), they must be
proportional to $e^2/l_0$, implying that $m^*\sim \sqrt{B}$.  
This would suggest that the gaps, measured in units of $e^2/\epsilon l_0$, are
proportional to $(2n+1)^{-1}$, consistent with the behavior found in our
calculations for the Coulomb interaction.  However, for the realistic gaps, the
effective mass has some filling factor dependence.
Fig.~\ref{fig8} shows the effective
mass determined from our theoretical gaps, along with the effective mass 
deduced from an analysis of  
the resistance oscillations at small $B^*$ in terms of  
Shubnikov-de Haas oscillations of ordinary fermions \cite {SdH,DuSdH}.
The experimental effective mass is seen to increase with $n$ \cite
{DuSdH,mass}; our results 
suggest that part of the increase may be caused by the short-distance
softening of the Coulomb interaction due to non-zero sample thickness.  
A logarithmic divergence of the mass predicted by the Chern-Simons
approach \cite {HLR} has a different physical origin; it 
is governed by the long-distance behavior of the interaction. 

\section{Model interactions}

Model interactions have been used in the past to study finite
thickness effects.  
There are other reasons for investigating how the gaps behave for 
various types of interaction.  First, certain analytical approaches find 
some forms of interaction more tractable,
and our Monte Carlo 
results provide a test for their validity \cite {Murthy}.
Second, the Chern-Simons field theoretical formulation of composite fermions
finds that the CF Fermi sea behaves qualitatively differently 
depending on whether the  interaction is of shorter or longer  
range than Coulomb \cite {HLR};  there are infrared singularities in the self
energy for the former,
indicating a divergent effective mass for composite fermions; for Coulomb
interaction, a logarithmic behavior is predicted, whereas no divergence 
occurs for interactions that are of longer range than Coulomb.  It is plausible
that some indication of this physics may be seen away from the CF sea, 
in the FQHE regime.  Finally, it 
may also be possible to actually change the form of the interaction, e.g., 
by fabricating the 2D electron gas close to a parallel conducting plane.
Motivated by these considerations, we have 
computed the gaps for various kinds of repulsive interactions: $1/r^2$;
logarithmic [$\ln 1/r$], Gaussian [$\exp(-r^2/2)$], Yukawa [$\exp(-r)/r$], 
and ZDS [$e^2/\sqrt{r^2+\lambda^2}$].   

\begin{figure}
\centerline{\psfig{figure=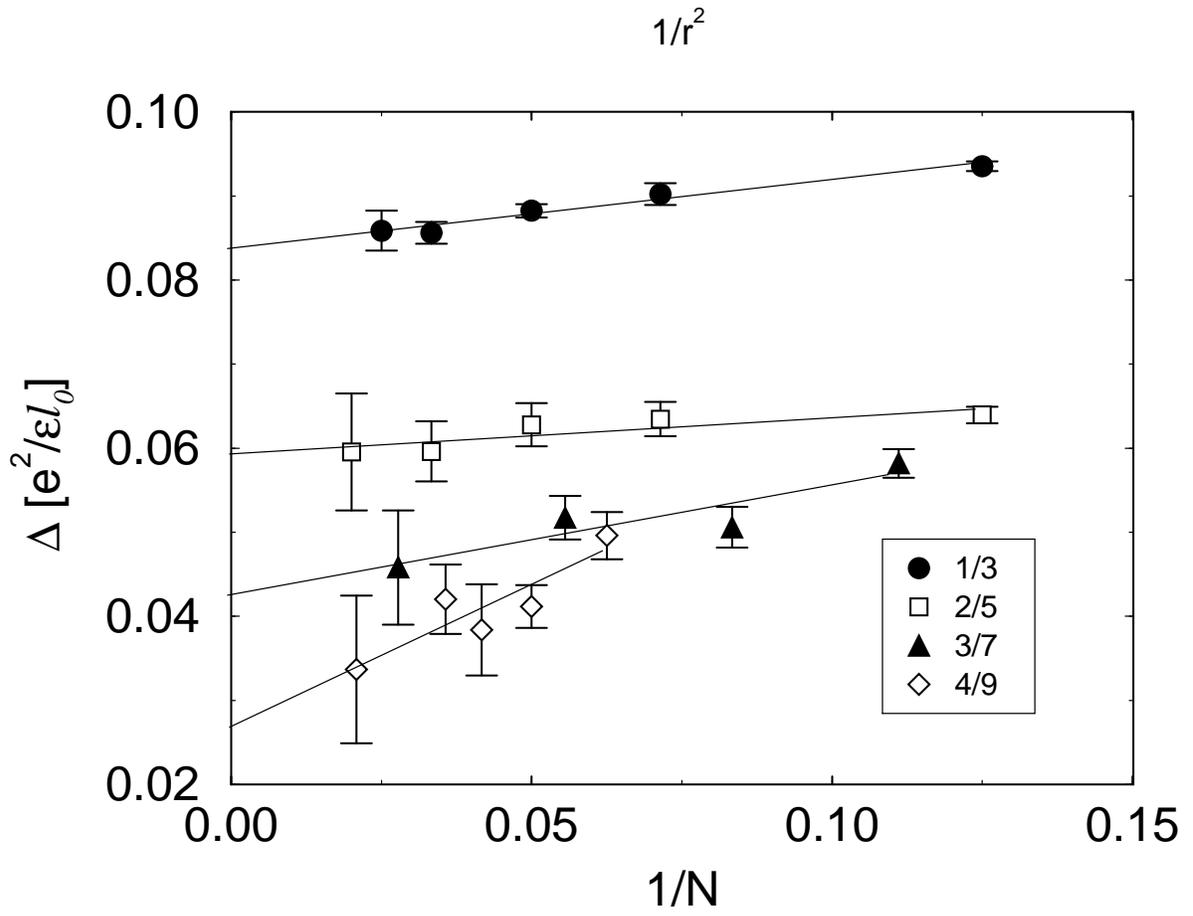,width=7.0in,angle=-90}}
\caption{Estimation of the thermodynamic limit of the gap from the finite
system results for the $r^{-2}$ interaction for 1/3, 2/5, 3/7, and 4/9. 
\label{fig9}}
\end{figure}

\begin{figure}
\centerline{\psfig{figure=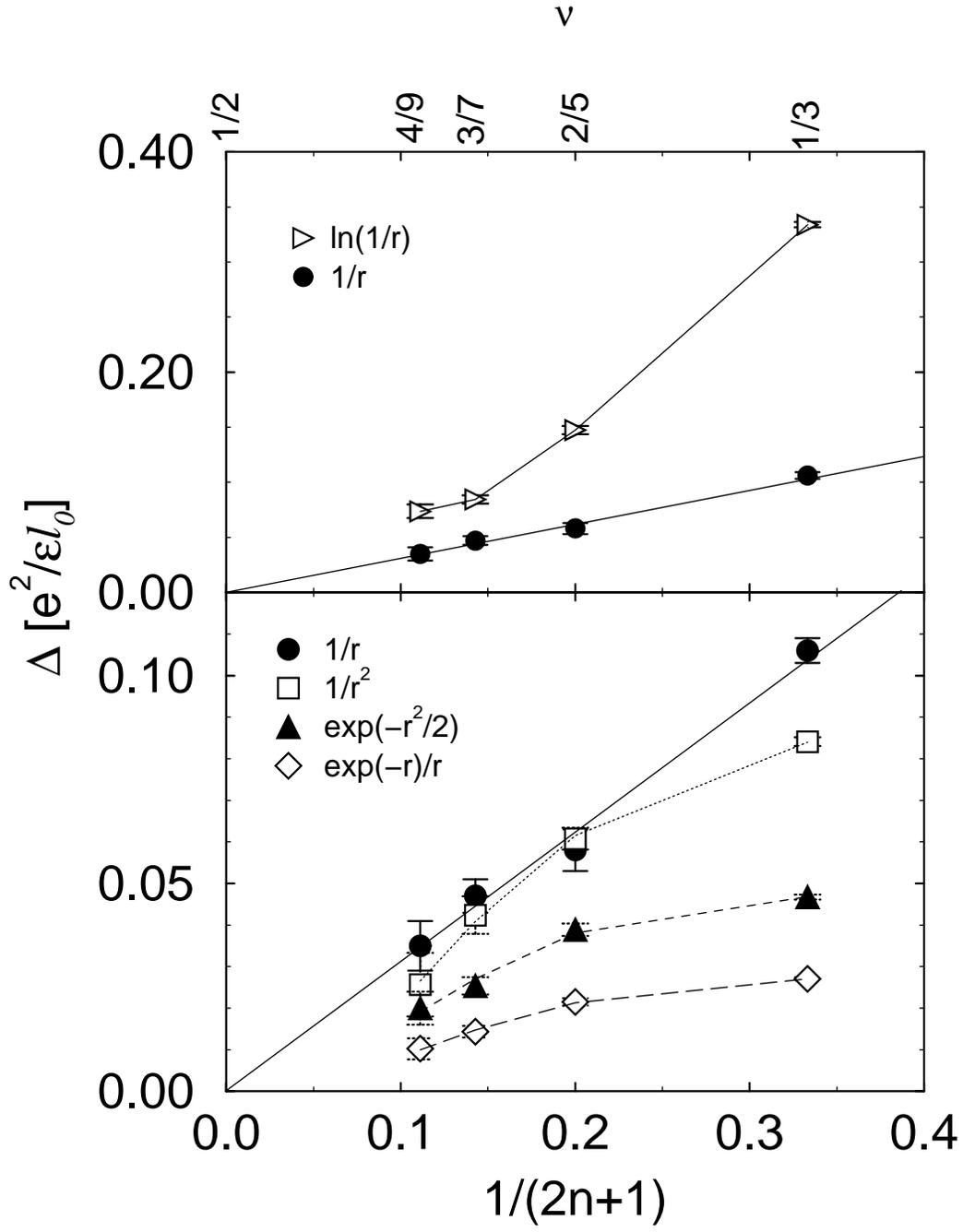,width=6.0in,angle=0}}
\caption{The activation gaps at 1/3, 2/5, 3/7, and 4/9
for several model interactions.  The pure Coulomb gaps are also
shown for reference.  All distances are quoted in  units of the 
magnetic length, $l_0$. 
\label{fig10}}
\end{figure}

The finite size extrapolation for the gaps are shown in Fig.~\ref{fig9}
 for the
$r^{-2}$ interaction.  Fig.~\ref{fig10} depicts the gaps for various
potentials;  the Coulomb results are included here for reference.  The longer
range potentials (e.g. logarithmic) have a qualitative different behavior
than the shorter range potentials.  In fact, there is indication that 
for the latter, the
gaps may vanish at a {\em finite} $n$, which we believe is related to 
the infrared divergences predicted by the Chern-Simons approach \cite {HLR}.
As stressed earlier, we 
are working with wave functions which are independent of the form
of the interaction, which raises the question of the relevancy of our 
study to the issue of stability of the CF sea; here, due to the 
lack of a gap to excitations, the wave functions, at in least their
long-distance behavior, will necessarily 
be highly susceptible to changes in the interaction.
We must remember, however, that the CF wave functions 
are expected to be accurate so long as the gap is not too
small, which is the case for the CF states with only a few filled CF-LLs.
Therefore, we believe that the trends seen in our study are meaningful.

\begin{figure}
\centerline{\psfig{figure=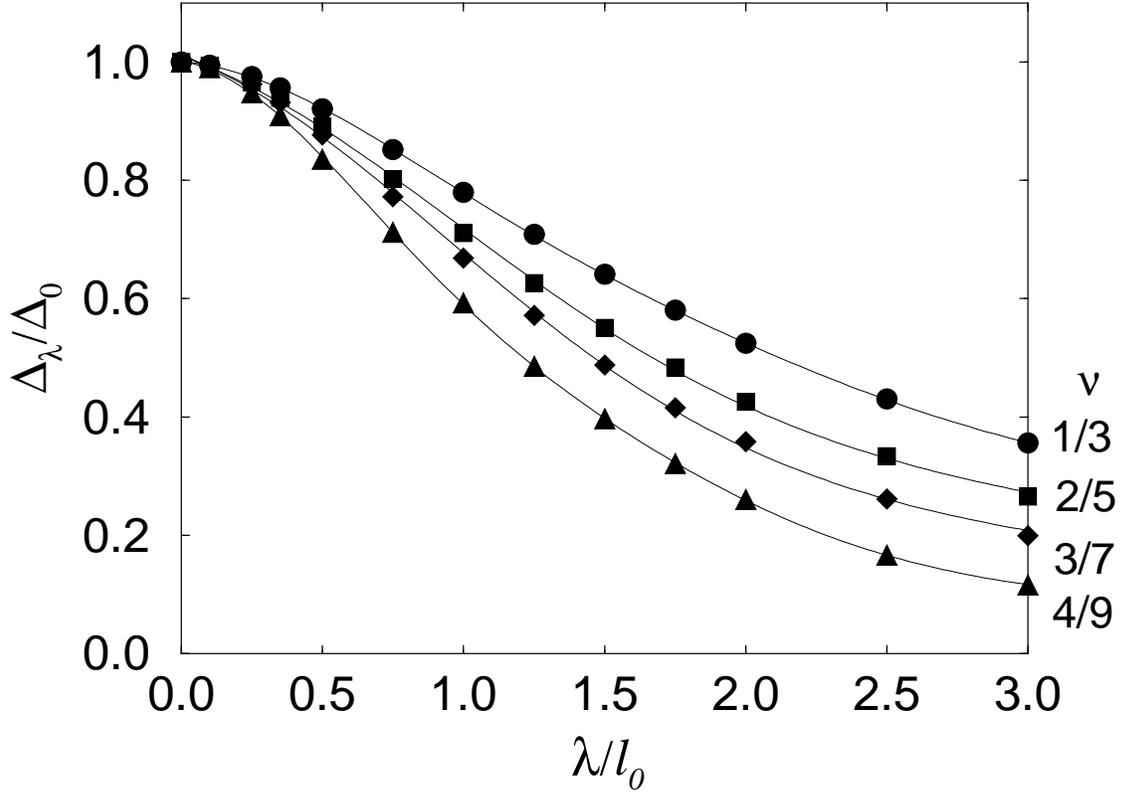,width=7.0in,angle=-90}}
\caption{The activation gaps at 1/3, 2/5, 3/7, and 4/9
for the Zhang Das Sarma potential, $e^2/(r^2+\lambda^2)^{1/2}$,
 plotted as a function of the parameter $\lambda$.  $\Delta_0$ is the gap for
pure Coulomb interaction. 
\label{fig11}}
\end{figure}

\begin{figure}
\centerline{\psfig{figure=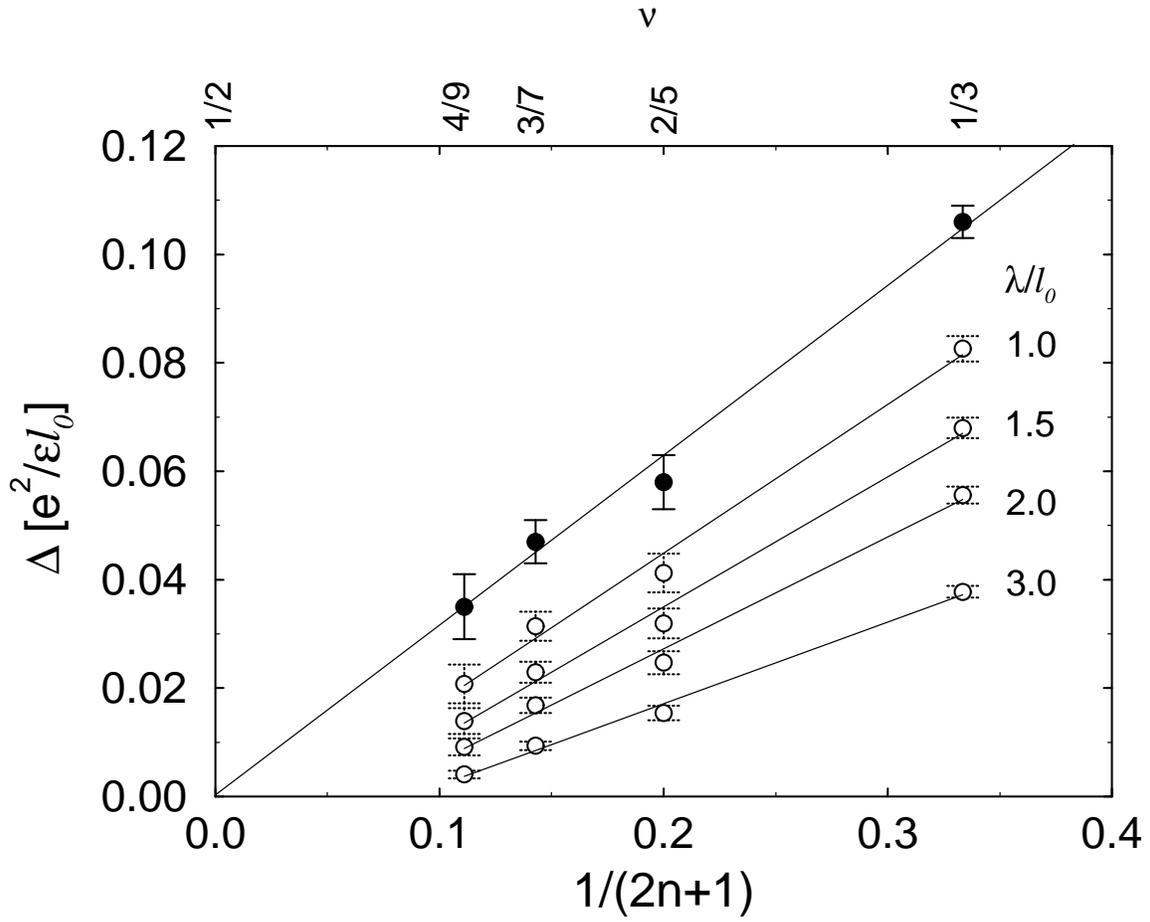,width=7.0in,angle=-90}}
\caption{The activation gaps for several values of $\lambda/l_0$ 
as a function of the filling factor 
for the Zhang Das Sarma potential.  The solid line is the best straight 
line fit
through the gaps.
\label{fig12}}
\end{figure}

\begin{figure}
\centerline{\psfig{figure=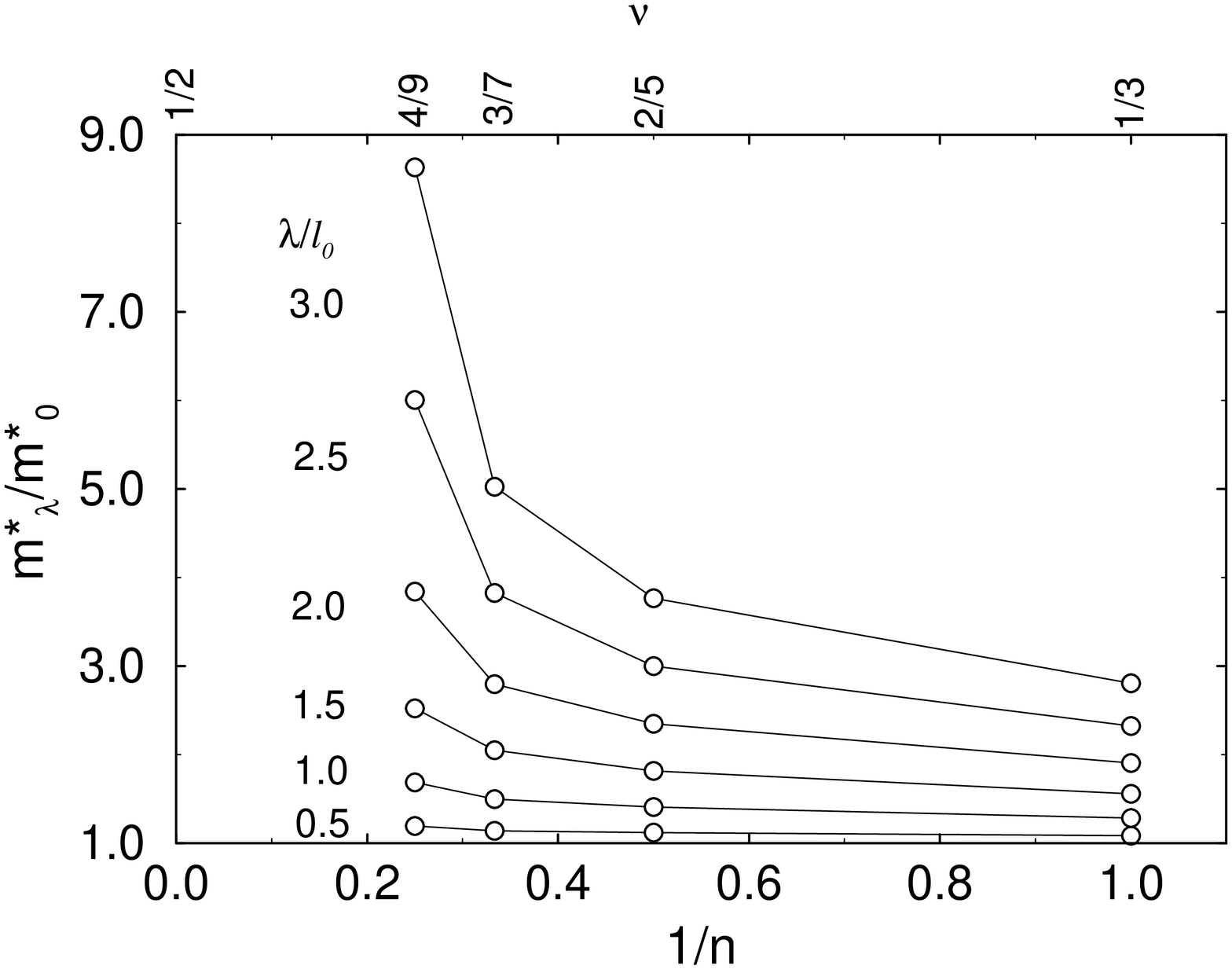,width=7.0in,angle=-90}}
\caption{The ratio of the CF effective mass for the ZDS interaction
(obtained from the gaps in 
Fig.~\ref{fig12}) to the effective mass for the Coulomb
interaction for several values of $\lambda/l_0$.
\label{fig13}}
\end{figure}

Fig.~\ref{fig11} shows the gaps for the ZDS potential as a function 
of $\lambda$.  The
gaps for fixed values of $\lambda/l_0$ are shown in Fig.~\ref{fig12}
and the effective
masses derived from them in Fig.~\ref{fig13}.  (Note that $\lambda/l_0$ is
kept fixed here rather than $\lambda$;  however, since the magnetic length 
does not change appreciably in the range of 
filling factors considered, the results are  
qualitatively independent of which of the two 
is taken as constant.)  These figures
demonstrate that for the ZDS potential also, 
similarly to the more realistic potentials, 
a straight line fit through the gaps
has a negative intercept, and the effective mass increases as the half filled
Landau level is approached.  The overall qualitative behavior is quite 
similar to that found in the more sophisticated LDA calculation;
a comparison of the two figures shows that the appropriate value for
$\lambda$ for the samples in the experiments of Du {\em et al.} \cite {Du} is 
$\lambda/l_0\approx 1$, as also argued by Morf \cite {Morf99}.   

\section{Conclusions}

We have carried out 
the most comprehensive study to date
of the effect of non-zero transverse width on 
activation gaps for the FQHE states.
The effective interaction between electrons has been computed  
by means of the density functional theory in the LDA, which is then 
used to determine the gaps for CF states with up to five filled CF-LLs
(corresponding to FQHE at 1/3, 2/5, 3/7, 4/9, and 5/11).  
Several different geometries are considered, and 
the theoretical results are compared to experiment.
It is concluded that for typical experimental parameters, 
the non-zero thickness reduces the gaps by 30\%, which does not 
fully account for the observed gaps.  This underscores the 
quantitative importance of effects left out in our study.

We have also considered a number of model interactions, and discovered a
qualitative difference depending on whether the interaction is of longer or
shorter range than Coulomb.  we find that the gaps for the FQHE states 
decrease faster for the latter, as the CF sea is approached, which is 
consistent with  expectations based on the Chern-Simons
formulation of composite Fermi sea \cite {HLR}, according to which 
the infrared behavior of the CF sea 
exhibits singularities for interactions of shorter range
than Coulomb but is well behaved for interactions of longer range 
than Coulomb.

It is a pleasure to thank S. Das Sarma and R. Morf for discussions.
This work was supported in part by the
National Science Foundation under grant no. DMR-9615005,
and by the National Center for Supercomputing Applications at the University
of Illinois (Origin 2000) under grant no. DMR970015N.

\pagebreak

\pagebreak

\pagebreak

\pagebreak

\pagebreak

\pagebreak

\pagebreak

\pagebreak

\pagebreak

\pagebreak

\pagebreak

\pagebreak

\pagebreak

\begin{thebibliography}{99}

\bibitem{Tsui} D.C. Tsui, H.L. Stormer, and A.C. Gossard,
Phys. Rev. Lett. {\bf 48}, 1559 (1982).

\bibitem{Laughlin} R.B. Laughlin, Phys. Rev. B {\bf 23}, 5632 (1981).

\bibitem{Jain}  J.K. Jain, Phys. Rev. Lett. {\bf 63}, 199 (1989);
Phys. Rev. B {\bf 41}, 7653 (1990); J.K. Jain and R.K. Kamilla 
in {\em Composite Fermions}, edited by Olle Heinonen (World Scientific, New
York, 1998). 

\bibitem{Review1} 
{\em Composite Fermions}, edited by Olle Heinonen (World Scientific, New York,
1998). 

\bibitem{Review2}
{\em Perspectives in Quantum Hall Effects}, edited by S. Das 
Sarma and A. Pinczuk (Wiley, New York, 1997).

\bibitem{tests} G. Dev and J.K. Jain, Phys. Rev. Lett. {\bf 69}, 2843 (1992);
X.G. Wu, G. Dev, and J.K. Jain, Phys. Rev. Lett. {\bf 71}, 153 (1993).

\bibitem{JK} J.K. Jain and R.K. Kamilla, Int. J. Mod. Phys. B {\bf 11}, 2621
(1997); Phys. Rev. B {55}, R4895 (1997).

\bibitem{Park} K. Park and J.K. Jain, Phys. Rev. Lett. {\bf 81}, 4200 (1998).

\bibitem{Morf} R. Morf and B.I. Halperin, Phys. Rev. B {\bf 33}, 2221
(1986).

\bibitem{Haldane} F.D.M. Haldane and E.H. Rezayi, Phys. Rev. Lett.  {\bf 54}, 237 (1985).

\bibitem{Fano} See, for example, 
G. Fano, F. Ortolani, and E. Colombo, Phys. Rev.  B {\bf 34}, 2670 (1986).

\bibitem{ZDS} F.C. Zhang and S. Das Sarma, Phys. Rev. B {\bf
33}, 2903 (1986).

\bibitem{Yoshioka} D. Yoshioka, J. Phys. Soc. Jpn. {\bf 55},
885 (1986).

\bibitem{Ortalano} M.W. Ortalano, S. He, and S. Das Sarma, 
Phys. Rev. B {\bf 55}, 7702 (1997).

\bibitem{Howard} F. Stern and W.E. Howard, Phys. Rev. {\bf 163}, 816.
 
\bibitem{Ando} T. Ando, A.B. Fowler, and F. Stern, Rev. Mod. Phys. {\bf 54},
437 (1982).

\bibitem{Morf99} R. Morf, Phys. Rev. Lett. {\bf 83}, 1485 (1999).

\bibitem{Park2}  A summary of some of our results appears in K. Park, N.
Meskini, and J.K. Jain, Phys. Rev. Lett. {\bf 83}, 1486 (1999).

\bibitem{LLmixing} D. Yoshioka, J. Phys. Soc. Jpn. {\bf 53},
3740 (1984); X. Zhu and S.G. Louie, Phys. Rev.
Lett. {\bf 70}, 339 (1993).

\bibitem{Stern} F. Stern and S. Das Sarma, Phys. Rev. B {\bf 30}, 840 (1984);
S. Das Sarma and F. Stern, {\em ibid.} B {\bf 32}, 8442 (1985).

\bibitem{HB} S.H. Vosko, L. Wilk, and M. Nusair,
Can. J. Phys. {\bf 58}, 1200 (1980).

\bibitem{Hedin} L. Hedin and B.I. Lundqvist, J. Phys. C {\bf 4}, 2064 (1971).

\bibitem{Willett} R.L. Willett {\em et al.}, 
Phys. Rev. B {\bf 37}, 8476 (1988).

\bibitem{Wu} T.T. Wu and C.N. Yang, Nucl. Phys. B {\bf 107}, 365
(1976); T.T. Wu and C.N. Yang, Phys. Rev. D {\bf 16}, 1018 (1977).

\bibitem{Haldane2} F.D.M. Haldane, Phys. Rev. Lett. {\bf 51}, 605 (1983).

\bibitem{Ceperley} D. Ceperley, G.V. Chester, M.H. Kalos; Phys. Rev. B
{\bf 16}, 3081 (1977); S. Fahy, X.W. Wang, and S.G. Louie,
Phys. Rev. B {\bf 42}, 3503 (1990).

\bibitem{Du} 
R.R. Du, H.L. Stormer, D.C. Tsui, L.N. Pfeiffer, and K.W.
West, Phys. Rev. Lett. {\bf 70}, 2944 (1993).

\bibitem{Manoharan} H.C. Manoharan, M. Shayegan, and S.J. Klepper,
Phys. Rev. Lett. {\bf 73}, 3270 (1994). 

\bibitem{Shayegan} M. Shayegan {\em et al.}, Phys. Rev. Lett. 
{\bf 65}, 2916 (1990).

\bibitem{HLR} B.I. Halperin, P.A. Lee, and N. Read, Phys. Rev. B {\bf
47}, 7312 (1993).

\bibitem{SdH} R.R. Du {\em et al.}, Solid State Commun. {\bf 90}, 71 (1994);
D.R. Leadley {\em et al.}, Phys. Rev.  Lett. {\bf 72}, 1906 (1994).

\bibitem{DuSdH} R.R. Du, H.L.
Stormer, D.C. Tsui, A.S. Yeh, L.N. Pfeiffer, and K.W. West, Phys. Rev.
Lett. {\bf 73}, 3274 (1994). 

\bibitem{mass}
For other experimental measurements of the CF effective mass, which are not 
all in agreement, see:
R.R. Du {\em et al.}, Phys. Rev. Lett. {\bf 70},, 2944 (1993);
P.T. Coleridge {\em et al.}, Phys. Rev. B {\bf 52}, R11603 (1995);
D.R. Leadley {\em et al.}, Phys. Rev. Lett. {\bf 72}, 1906 (1994).

\bibitem{Murthy} See, for example, 
G. Murthy {\em et al.}, Phys. Rev. B {\bf 58} 15363, 1998. 


\end{thebibliography}
\end{document}